\documentclass[useAMS,usenatbib,letters] {mn2e}
\usepackage{psfig}
\usepackage{graphicx}
\usepackage{amssymb} 

\title[Arches Stellar Cluster Mass Function]{The origin of the Arches stellar cluster mass function}

\author[Dib, Kim, \& Shadmehri]{Sami Dib$^{1}$\thanks{E-mail: dib@kasi.re.kr}, Jongsoo Kim$^{1}$, and  Mohsen Shadmehri$^{2,3}$\\
$^{1}$Korea Astronomy and Space Science Institute, 61-1, Hwaam-dong, Yuseong-gu, Daejeon 305-348, Korea\\
$^{2}$School of Mathematical Sciences, Dublin City University, Glasnevin, Dublin 9, Ireland\\
$^{3}$Department of Physics, School of Science, Ferdowsi University, Mashad, Iran}

\begin{document}
\maketitle

\date{Accepted XXX. Received XXX}

\pagerange{\pageref{firstpage}--\pageref{lastpage}}
\pubyear{2007}
\label{firstpage}

\begin{abstract} 
We investigate the time evolution of the mass distribution of pre-stellar cores (PSCs) and their transition to the initial stellar mass function (IMF) in the central parts of a molecular cloud (MC) under the assumption that the coalescence of cores is important. Our aim is to explain the observed shallow IMF in dense stellar clusters such as the Arches cluster. The initial distributions of PSCs at various distances from the MC center are those of gravitationally unstable cores resulting from the gravo-turbulent fragmentation of the MC. As time evolves, there is a competition between the PSCs rates of coalescence and collapse. Whenever the local rate of collapse is larger than the rate of coalescence in a given mass bin, cores are collapsed into stars. With appropriate parameters, we find that the coalescence-collapse model reproduces very well all the observed characteristics of the Arches stellar cluster IMF; Namely, the slopes at high and low mass ends and the peculiar bump observed at $\sim 5-6$ $M_{\odot}$. Our results suggest that today's IMF of the Arches cluster is very similar to the primordial one and is prior to the dynamical effects of mass segregation becoming important.
\end{abstract} 


\section{MOTIVATION}

Understanding the origin of the initial stellar mass function (IMF) remains one of the most challenging issues in modern astrophysics. When averaged over the total volume of galaxies or whole stellar clusters, the IMF is observed to follow a nearly uniform behavior which consists in an increased number of stars counted when going from the most massive stars up to $\sim 0.5$ $M_{\odot}$, followed by a shallower increase between $\sim 0.5$ and $\sim 0.1$ $M_{\odot}$ and a decline in the number of stars at masses $\lesssim 0.1$ $M_\odot$. This standard IMF has been described, with continuous refinements, by several analytical functions (e.g., Salpeter 1955; Miller-Scalo 1979; Kroupa 2002; Chabrier 2003). Yet, deviations from the standard IMF at low and high mass ends have been reported in many observations (see review in Elmegreen 2004). At high mass, the IMF is observed to be generally top-heavy in dense cluster cores such as in the Arches cluster (e.g., Stolte et al. 2005; Kim et al. 2006) and stars appear to be, preferentially located in the central parts of the clusters (e.g., Hillenbrand \& Hartmann 1998; Figer et al. 1999; Stolte 2002; Gouliermis et al. 2004). Star-bursts regions are also observed to possess a top-heavy IMF, either in the form of a shallow slope at high mass (e.g., Einsenhauer et al. 1998; Sternberg 1998) or by having a value of the high mass-low mass turnover of a few to several $M_\odot$ which is substantially larger than that of the standard IMF (e.g., Rieke 1993). The IMF of dense clusters seems also to be truncated at the very high mass end (e.g., Stolte 2005).  

The mass truncation can be attributed to the short lifetimes of the most massive stars. Ideas that have been proposed to explain the shallowness of the slope at the high mass end include a) a model based on the coalescence of pre-stellar cores (PSCs) and their subsequent gravitational collapse to produce stars (e.g., Nakano 1966; Silk \& Takahashi 1979; Elmegreen \& Shadmehri 2003; Elmegreen 2004; Shadmehri 2004), b) the mass segregation of stars in the cluster (e.g., Vesperini \& Heggie 1997; Kroupa 2002; Mouri \& Taniguchi 2002), and c) A renewed episode of gas accretion by the cluster under favorable conditions, which leads to the formation of a new generation of massive stars (e.g., Lin \& Murray 2007). This latter idea is somehow inconsistent with the fact that a cluster such as the Arches cluster is overall very young (i.e, age $\sim 2 \pm 1$ Myrs) and may apply only to older clusters. Concerning mass segregation, whereas there is little doubt that the enhancement in the numbers of massive stars in the inner parts of the cluster by dynamical processes will lead to a shallower IMF, this does not constitute a direct proof that the primordial IMF of stars in those regions was not shallower than a Salpeter IMF initially. The latter is commonly used as an initial input for the stellar distribution functions at all cluster radii in N-Body models (e.g., Portegies Zwart et al. 2007). Furthermore, the IMF of the Arches cluster is characterized by a peculiar bump at $\sim 6$ $M_{\odot}$ which is not, to date, well reproduced by the effect of mass segregation in N-body simulations (e.g., Kim et al. 2006, Portegies Zwart et al. 2007). 

In this letter, we propose a coalescence model in which the local initial PSCs populations are those resulting from the local gravo-turbulent fragmentation of the protocluster cloud. We follow the time evolution of the mass function of PSCs and the transition to the IMF under the assumption that the coalescence of PSCs is important. This is very likely to be the case for the PSCs located in the central parts of the protocluster cloud. 

\section{THE COALESCENCE MODEL}\label{model}

We consider PSCs (e.g., Andr\'{e} et al. 2000) embedded in an isothermal MC (at a temperature of $T=10$ K), at different locations $r$ from the cloud's center. We assume that both the PSCs and the MC are axisymmetric (PSCs are initially spherical but are likely to quickly flatten as time evolves). The radial density profile of the MC is given by: 

\begin{equation} 
\rho_{c}(r)= \frac{\rho_{c0}}{1+(r/R_{c0})^{2}},
\label{eq1}
\end{equation}
 
where $\rho_{c0}$ and $R_{c0}$ are the cloud's central density and core radius, respectively. The central density, $\rho_{c0}$, is given by:

\begin{equation} 
\rho_{c0}= \frac{M_{c}}{4 \pi R_{c0}^{3} [(R_{c}/R_{c0})-arctan(R_{c}/R_{c0})]},
\label{eq2}
\end{equation}

where $M_{c}$ is the mass of the cloud and $R_{c}$ its radius. The density profiles of PSCs are assumed to follow the formula given by Whitworth \& Ward-Thompson (2001): 

\begin{equation} 
\rho_{p}(r_{p})= \frac{\rho_{p0}}{[1+(r_{p}/R_{p0})^{2}]^{2}},
\label{eq3}
\end{equation}

where $\rho_{p0}$ and $R_{p0}$ are the central density and core radius of the PSC, respectively. The radius $R_{p}$ of the the PSC depends both on its mass and on its position within the MC. The dependence of $R_{p}$ on $r$ requires that the density at the edges of the PSC equals the ambient cloud density, i.e., $\rho_{p}(R_{p})=\rho_{c}(r)$. This would result in smaller radii for PSCs of a given mass when they are located in their inner parts of the cloud. The density contrast between the edge of the PSC and its center is given by 

\begin{equation} 
{\cal C}(r) = \frac {\rho_{p0}}{\rho_{c} (r)}=\frac {\rho_{p0}} {\rho_{c0}} \left(1+ \frac{r^{2}}{R^{2}_{c0}} \right).         
\label{eq4}
\end{equation}

Depending on its position $r$ in the cloud, the radius of the PSC of mass $M_{p}$, $R_{p}$, can be calculated as being $R_{p}(r,M_{p})=a(r) R_{p0} (r,M_{p})$, where: 

\begin{equation} 
R_{p0}(r,M_{p})= \left(\frac{M_{p}}{2 \pi \rho_{p0}} \right)^{1/3} \left(arctan[a(r)]-\frac{a(r)}{1+a(r)^{2}} \right)^{-1/3},
\label{eq5}
\end{equation} 

and with $a(r)=({\cal C}(r)^{1/2}-1)^{1/2}$. With our set of parameters, the quantity ${\cal C}^{1/2}-1$ is always guaranteed to be positive. The value $R_p(r,M)$ can be considered as being the radius of the PSC at the moment of its formation. However, the radius of the PSC will decrease as time advances due to gravitational contraction. The PSC contracts on a timescale, $t_{cont,p}$ which is equal to a few times its free fall timescale, and can be parametrized as:

\begin{equation} 
t_{cont,p}(r,M)= \nu ~ t_{ff}(r,M)= \nu \left( \frac {3 \pi} {32~G \bar{\rho_{p}} (r,M)} \right)^{1/2},
\label{eq6}
\end{equation}

where $\nu \ge 1$ and $\bar{\rho_{p}}$ is the radially averaged density of the PSC of mass $M_{p}$, located at position $r$ in the cloud, and which is calculated as being:

\begin{equation} 
\bar{\rho_{p}}(r,M_{p})=\frac{1}{R_{p}(r,M_{p})}~\int^{R_{p}(r,M_{p})}_{0}\frac{\rho_{p0}}{[1+(r_{p}/R_{p0})^{2}]^{2}} dr_{p},
\label{eq7}
\end{equation}

Thus, the time evolution of the radius of the PSC can be described by the following equation: 

\begin{equation} 
R_{p}(r,M,t)=R_p(r,M)~e^{-(t/t_{cont,p})}.
\label{eq8}
\end{equation}

Once the instantaneous radius of a PSC of mass $M_{p}$, located at position $r$ form the cloud's center is defined, it becomes possible to calculate its cross section for collision with PSCs of different masses. The cross section for the collision of a PSC of mass $M_{i}$ and radius $R_{i}$ with another of mass $M_{j}$ and radius $R_{j}$ and which accounts for the effect of gravitational focusing is given by: 

\begin{eqnarray} 
\sigma(M_{i},M_{j},r,t) = \pi \left[R_{p,i}(r,M_{i},t)+R_{p,j}(r,M_{j},t)\right]^{2} \nonumber \\
\times \left[ 1+\frac{2G (M_{i}+M_{j})} {2 v^{2} (R_{p,i}(r,M_{i},t)+R_{p,j}(r,M_{j},t))} \right].
\label{eq9}
\end{eqnarray}

Elmegreen \& Shadmehri (2003) and Shadmehri (2004) assumed that the collision velocity between PSCs is equal to the virialized velocity dispersion inside the MC. This might be a plausible hypothesis if MCs were indeed the dissipative structures of turbulence in the interstellar medium. It is however unlikely to be the case. Numerical simulations (e.g., Dib et al. 2007) show that clumps and cores in MCs are not in virial equilibrium. In this work, we assume that the relative collision velocity between the PSCs follows the local gas dynamics at their position in the cloud (this remains a simplification as in reality PSCs motions can be decoupled from that of the local ambient gas) according to a Larson type relation $v(r)=v_{0} r^{\alpha}$ (Larson 1981; $v_{0}=1.1$ km s$^{-1}$), with a lower limit being the local thermal sound speed, which is uniform across the isothermal MC. 

\section{INITIAL CONDITIONS} 

As initial conditions for the PSCs mass distribution at different cloud radii, we adopt distributions that are the result of the gravo-turbulent fragmentation of the cloud, following the formulation given in Padoan et al. (1997) and Padoan \& Nordlund (2002). In these models, the probability function of the density field is well represented by a log-normal function:

\begin{equation}
P(ln~x) d~ln~x=\frac{1}{\sqrt{2 \pi \sigma_{d}}} exp \left[-\frac{1}{2} \left(\frac{ln~x-\bar{ln~x}}{\sigma_{d}} \right)^{2} \right] d~ln~x,
\label{eq10}
\end{equation}

where $x$ is the number density normalized by the average number density, $x=n/\bar{n}$. The standard deviation of the density distribution $\sigma_{d}$ and the mean value $\bar {ln~x}$ are functions of the thermal rms Mach number, $\cal M$: $\bar{ln x}=-\sigma^{2}_{d}/2$ and $\sigma^{2}_{d}=ln(1+{\cal M}^{2} \gamma^{2})$. Padoan \& Nordlund (2002) suggest a value of $\gamma \sim 0.5$. A second step in this approach is to determine the mass distribution of dense cores. Padoan \& Nordlund (2002) showed that by making the assumptions that: (a) the power spectrum of turbulence is a power law and, (b) the typical size of a dense core scales as the thickness of the postschock gas layer, the cores mass spectrum is given by:

\begin{equation}
N(M)~d~log~M \propto M^{-3/(4-\beta)} d~log~M,
\label{eq11}
\end{equation}   

where $\beta$ is the exponent of the kinetic energy power spectrum, $E_{k} \propto k^{-\beta}$, and is related to the exponent $\alpha$ of the size-velocity dispersion relation in the cloud with $\beta=2 \alpha+1$. However, Eq.~\ref{eq11} can not be directly used to estimate the number of cores that are prone to star formation. It must be multiplied by the local distribution of Jeans masses. At constant temperature, this distribution can be written as:

\begin{equation}
P(M_{J})~dM_{J}=\frac{2~M_{J0}^{2}}{\sqrt{2 \pi \sigma^{2}_{d}}} M^{-3}_{J} exp \left[-\frac{1}{2} \left(\frac{ln~M_{J}-A}{\sigma_{d}} \right)^{2} \right] dM_{J},
\label{eq12}
\end{equation}   

where $M_{J0}$ is the Jeans mass at the mean density $\bar{n}$. Thus, Eq.~\ref{eq11} becomes, locally: 

\begin{equation}
N (r,M)~d~log~M =f_{0}(r)~M^{-3/(4-\beta)} \left[\int^{m}_{0} P(M_{J}) dM_{J}\right]d~log~M.
\label{eq13}
\end{equation}   

The local normalization coefficient $f_{0}(r)$ is obtained by requiring that $\int^{M_{max}}_{M_{min}} N (r,M)~dM=1$ in the shell of width $dr$ located at distance $r$ from the cloud's center. Then, the local distribution of cores at time $t=0$, $N(r,M,0)$, is obtained by multiplying the local normalized function $N(r,M)$ by the local rate of fragmentation such that:

\begin{equation}
N(r,M,0)=\frac{\epsilon_{c}(r) \rho_{c}(r)} {<M>(r)~t_{cont,p} (r,M)} N(r,M),
\label{eq14}
\end{equation}

where $<M>$ is the average core mass in the local distribution and is calculated by $<M>=\int_{M_{min}}^{M_{max}} M~N(r,M)~dM$ and $\epsilon_{c}$ is a parameter smaller than unity which describes the local mass fraction of gas that is present in the dense PSCs. In principle, $\epsilon_{c}$ will have a radial and probably outwardly decreasing dependence. For simplicity we shall assume $\epsilon_{c}$ to be a constant independent of radius. As our comparisons with the observations will be focused on the inner parts of the protocluster cloud which will be transformed into a stellar cluster (i.e., the Arches cluster), it is likely that these regions will be characterized by a uniform mass fraction of the dense gas. 
 
\begin{figure}
\begin{center}
\psfig{figure=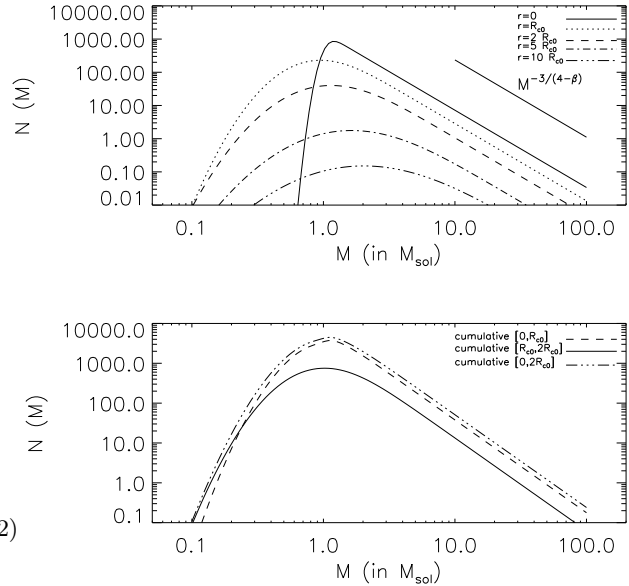,width=\columnwidth}
\end{center}
\caption{Top: Mass spectrum of Jeans unstable pre-stellar cores in shells of width 0.025 pc located at different distances from the cloud center (at 0, 1, 2, 5, and 10 $R_{c0}$), where $\beta$ is the exponent of the kinetic energy power spectrum. Bottom: Cumulative number of cores in the regions between [0,$R_{c0}$], [$0,2~R_{c0}$], and [$R_{c0},2~R_{c0}$].}
\label{fig1}
\end{figure}

Fig.~\ref{fig1} displays the local mass spectrum of Jeans unstable PSCs in rings of width 0.025 pc, obtained with Eq.~\ref{eq14}, located at different distances from the cloud's center (top), as well as the cumulative number of PSCs in each mass bin in regions of the protocluster cloud located between [$0,R_{c0}$], [$0,2~R_{c0}$], and [$R_{c0},2~R_{c0}$].

\section{FROM THE PRE-STELLAR CORES MASS FUNCTION TO THE PRIMORDIAL IMF}

\begin{figure*}
\begin{center}
\includegraphics[height=12cm, width=17.5cm]{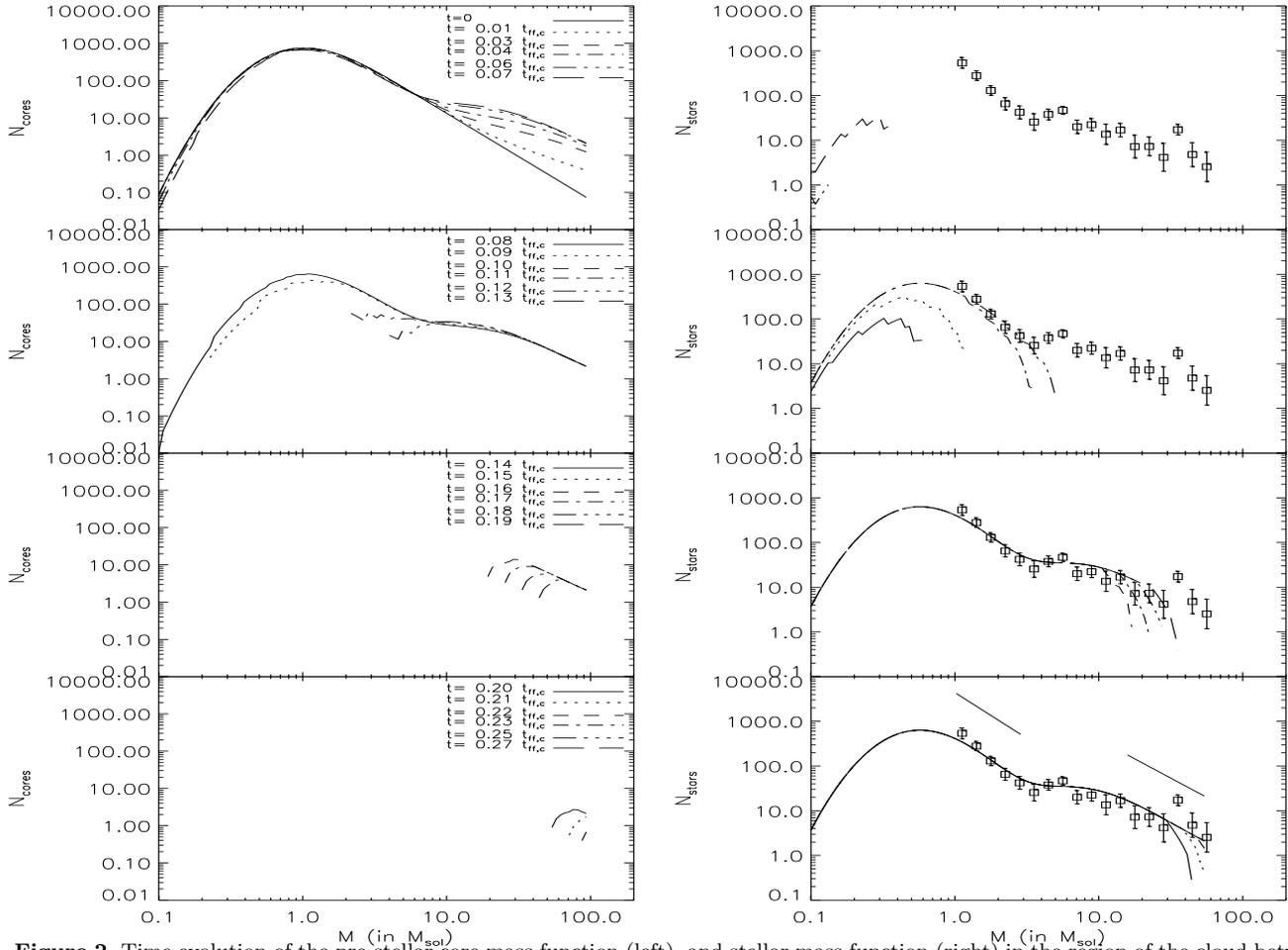}
\end{center}
\caption{Time evolution of the pre-stellar core mass function (left), and stellar mass function (right) in the region of the cloud between [$R_{c0},2~R_{c0}$]. The stellar mass function is compared to that of the Arches stellar cluster mass function (Kim et al. 2006). Fits to the simulated IMF (bottom right figure) yield slopes of $-2.04 \pm 0.02$ and $-1.72 \pm 0.01$ in the mass ranges of [$1-3$] $M_{\odot}$ and $\ge 15$ $M_{\odot}$, respectively, in very good agreement with the observations. Fits are over-plotted to the data shifted up by 1 dex for the sake of clarity.}
\label{fig2}
\end{figure*} 

With the initial conditions described in \S.~3, we follow the time evolution of the PSCs mass spectrum by solving the following integro-differential equation of $N(r,M,t)$:

\begin{eqnarray} 
\frac{dN(r,M,t)}{dt}=0.5\times \eta(r) \times \nonumber \\
 \int^{\Delta M}_{M_{min}}~N(r,m,t)~N(r,M-m,t)~\sigma(m,M-m,r,t)~v(r)~dm \nonumber \\ 
 -\eta (r) N(r,M,t) \int^{M_{max}}_{M_{min}} N(r,m,t) \sigma(m,M,r,t) v(r)~dm                            
\label{eq15}
\end{eqnarray}

where the first and second terms in the right hand side of Eq.~\ref{eq15} correspond to the rate of creation and destruction of a PSC of mass $M$, at location $r$, respectively (Nakano 1966; Shadmehri 2004). In Eq.~\ref{eq15}, $\Delta M=M-M_{min}$, and $\eta (r)$ is a coefficient which represents the coalescence efficiency, with $\eta \leq 1 $. This efficiency can be the result of various physical processes which can affect the coalescence of PSCs, such as if the merger of cores occurs preferentially parallel or perpendicular to the local magnetic field lines, and is likely to have a radial dependence. For simplicity, we shall assume that $\eta$ is independent of position. In order to evaluate the transition from PSCs to stars, we compare, at each timestep, the local coalescence timescale to the local contraction timescale for PSCs of a given mass. The local coalescence timescale is $t_{coal}(r,M)=1/w_{coal}(r,M)$ where $w_{coal}$ is the coalescence rate (Elmegreen \& Shadmehri 2003): 

\begin{equation}
w_{coal}(r,M)=\frac{2^{1/2} v(r)}{V_{shell}(r)} \sum_{j=1}^{mbin} (R_{i}+R_{j})^{2} \left[1+\frac{2 G (M_{i}+M_{j})}{2 v^{2} (R_{i}+R_{j})} \right],
\label{eq16}
\end{equation}   

where $mbin$ is the number of mass bins, and $V_{shell}$ is the volume of the shell of width $dr$ located at distance $r$ from the MC's center. The contraction timescale is given by Eq.~\ref{eq8}. Whenever the local contraction timescale is shorter than the local coalescence timescale, PSCs are collapsed into stars. When a PSC collapses to form a star, we assume that a fraction of its mass is re-injected into the protocluster cloud in the form of an outflow. We account for this mass loss in a purely phenomenological way by assuming that the mass of a star which is formed out of a PSC of mass $M_{p}$ is given by M$_{\star}$=$\psi$ M$_{p}$, where $\psi \le 1$. Matzner \& McKee (2000) showed that $\psi$ can vary between $0.25-0.7$ for stars in the mass range $0.5-2$ $M_{\odot}$. There is no evidence so far, for or against, whether this result holds at higher masses. However, the similarity between the IMF and the dense cores mass function observed by Alves et al. (2007) in the Pipe Nebula might be an indication of a constant $\psi$ across the mass spectrum (i.e., in their case it is $\psi \sim 1/3$). Here also, we shall assume that a similar fraction of the mass of a PSC is lost in the outflow independent of its mass. 

The algorithm was tested by performing runs with $\eta=0$ (i.e., no-coalescence) and $\eta=0.001$ (i.e., inefficient coalescence) and with the other parameters fixed at $M_{c}=5 \times 10^{5}$ $M_{\odot}$, $R_{c}=5$ pc, $R_{c0}=0.2$ pc, $\rho_{p0}=10^{7}$ cm$^{-3}$, $\epsilon=0.5$, $\alpha=0.37$, $\nu=10$, and $\psi=0.58$. As expected, for $\eta=0$, the resulting stellar mass spectrum after the PSCs collapse into stars is similar to the initial cumulative PSCs spectrum, and only slightly modified for $\eta=0.001$. Models were performed with permutations over the parameters $\eta$ and $\nu$ fixing the other parameters to the above stated values. It should be stressed at this stage that our semi-analytical modeling is not aimed at recovering the initial characteristics of the Arches protocluster cloud, but rather at showing whether or not, the Archer cluster IMF can be generated by the coalescence of PSCs and their subsequent collapse into stars.

 Fig.~\ref{fig2} displays the time evolution of the cumulative PSCs populations in the region [$R_{c0}-2~R_{c0}$]=[0.2-0.4] pc, which corresponds to the annulus between $\sim$ 1-2 core radii of the Arches cluster for a model with $\eta=0.5$ and $\nu=10$. In the initial stages, the most massive PSCs, who have a larger cross section, coalesce faster than the less massive ones, essentially by capturing the numerous intermediate mass PSCs and causing a rapid flattening of the spectrum at the high mass end. By $t \sim 0.07~t_{ff,c}$ ($t_{ff,c}=(3 \pi/32 G \bar{\rho_{c}})^{1/2} \sim 3 \times 10^{4}$ yr is the MC free fall timescale), a first generation of the smallest PSCs collapses to form stars. As time advances, more massive stars are formed in the shell (massive cores collapse later because of their lower average density) and in parallel the PSCs population decreases. By $t \sim 0.1~t_{ff,c}$ the intermediate mass PSCs which constitutes the largest mass reservoir for coalescence collapse into stars. At this time, the turnover in the PSCs mass spectrum is located at $\sim 8-10$ $M_{\odot}$. Since the reservoir of intermediate mass objects is depleted, the remaining massive PSCs coalesce at a slower pace before they collapse. By $t \sim 0.25~t_{ff,c}$, all PSCs of different masses in the shell have collapsed and formed stars. Because of mass loss, the stellar IMF is shifted to lower masses (bump shifted to $\sim 5-6$ $M_{\odot}$). In summary, the resulting IMF is not very different from the PSCs mass spectrum after the initial and rapid stage of strong coalescence until $t \sim 0.01~t_{ff,c}$. This is due to the fact that low and intermediate mass PSCs collapse at early stages, thus depleting the reservoir of objects with which the massive PSCs can continue to coalesce, in addition to their own contraction. Both effects reduce the massive PSCs ulterior merger rate. Overall, the stellar mass spectrum is formed very quickly, on a timescale which is of the order of the contraction timescale of the most massive cores i.e., $\sim 0.25~t_{ff,c}$.

 In Fig.~\ref{fig2}, over-plotted to our result is the cumulative mass spectrum of the Arches cluster in the annulus of [0.2-0.4] pc (Kim et al. 2006). The coalescence-collapse model agrees better with the observations than models based on mass segregation by dynamical friction. In particular, the bump at $\sim 5-6$ $M_{\odot}$ is reproduced. A fit to the stellar spectrum yields slopes of $-2.04 \pm 0.02$ and $-1.72 \pm 0.01$ in the mass ranges of [$1-3$] $M_{\odot}$ and $\ge 15$ $M_{\odot}$, respectively, in very good agreement with observational values. 

We also performed additional runs where the maximum mass in the PSC spectrum was set to 250 $M_{\odot}$ (instead of 100 $M_{\odot}$). In this case, the resulting slope of the IMF in the low and high mass regimes are shallower than the Salpeter IMF, yet shallower than those of the Arches IMF. The reason is that PSCs with masses larger than $100$ $M_{\odot}$ will form quickly from the coalescence of lower mass ones, and the number of PSCs of masses $\gtrsim 100$ $M_{\odot}$ will grow at an even faster pace as their cross sections are very large. The mismatch in this case with the Arches IMF might be an indication that PSCs with masses $\ge 100$ $M_\odot$, if they form, might undergo a certain amount of sub-fragmentation.      

\section{SUMMARY}

In this work, we use semi-analytical modeling to study the evolution of the pre-stellar cores (PSCs) mass spectrum and its transition to the stellar initial mass function (IMF) at different locations in a molecular cloud (MC) under the assumption that the coalescence of PSCs is important. The aim is to reproduce the observed IMF in the inner regions of dense stellar clusters such as the Arches cluster (Kim et al. 2006). The initial conditions for the local populations of PSCs are those of Jeans unstable cores resulting from the gravo-turbulent fragmentation of the MC. PSCs of a given mass are transformed into stars whenever their local rate of contraction is higher than their rate of coalescence. With appropriate, yet very realistic parameters, we are able to reproduce all of the observed characteristics of the IMF of the Arches cluster. Namely, the slopes at the high and low mass ends, and the peculiar bump observed at $\sim 5-6$ $M_{\odot}$. Our results suggest that today's IMF of the Arches cluster is primordial. This might be a common property of young and dense stellar clusters (e.g., Chen et al. 2007). Another consequence of the coalescence-collapse model is that it might help explain the formation of intermediate-mass black holes ($M_{BH} \gtrsim  100$ $M_{\odot}$) in the central regions of dense stellar clusters, either by the direct gravitational collapse of massive PSCs or by the runaway collisions of massive stars (e.g., Bonnell et al. 1998; Freitag et al. 2006) which would be fostered if the primordial IMF is top-heavy.   
       
\section*{Acknowledgments}
We thank the anonymous referee for valuable comments and Dongsu Ryu, Christopher Matzner, Carl Jakob Walcher, Sungsoo Kim, and Zhi-Yun Li for useful discussions.
 
{}

\label{lastpage}


\begin{thebibliography}{}

\bibitem[Alves (2007)] {alves07} Alves, J., Lombardi, M., \& Lada, C. J. 2007, A\&A, 462, L17  
\bibitem[Andre (2000)] {andre00} Andr\'{e}, P., Ward-Thompson, D., \& Barsony, M.  2000, in Protostars and Planets IV,   ed. V. Mannings, A. P. Boss, S. S. Russel (Tuscon:University of Arizona Press), 59 
\bibitem[Bonnell (1998)] {bonnell98} Bonnell, I. A., Bate, M. R., \& Zinnecker, H. 1998, MNRAS, 298, 93 
\bibitem[Chabrier (2003)] {chabrier03} Chabrier, G. 2003, PASP, 115, 763 
\bibitem[Chen (2007)] {chen07} Chen, L., de Grijs, R., \& Zhao, J. L. 2007, AJ, accepted, (arXiv:0706.2723)
\bibitem[Dib (2007)] {dib07} Dib, S., Kim, J., V\'{a}zquez-Semadeni, E., Burkert, A., \& Shadmehri, M. \ 2007, ApJ, 661, 262 
\bibitem[Eisenhauer (1998)] {eisenhauer98} Eisenhauer, F., Quirrenbach, A., Zinnecker, H., \& Genzel, R. 1998, ApJ, 498, 278 
\bibitem[Elmegreen (2003)] {elmegreen03} Elmegreen, B. G., \& Shadmehri, M. 2003, MNRAS, 338, 817
\bibitem[Elmegreen (2004)] {elmegreen04} Elmegreen, B. G. 2004, MNRAS, 354, 367 
\bibitem[Figer (1999)] {figer99} Figer, D. F., McLean, I. S., \& Morris, M. 1999, ApJ, 514, 202
\bibitem[Freitag (2006)] {freitag06} Freitag, M., Atakan G\"{u}rkan, M., \& Rasio, F. 2006, MNRAS, 368, 141 
\bibitem[Gouliermis (2004)] {gouliermis04} Gouliermis, D., Keller, S. C., Kontizas, M., Kontizas, E., \& Bellas-Velidis, I.  2004, A\&A, 416, 137
\bibitem[Hillenbrand (1998)] {hillenbrand98} Hillenbrand, L. A., \& Hartmann, L. W.  1998, ApJ, 492, 540
\bibitem[Kim (2006)] {kim06} Kim, S. S., Figer, D. F., Kudritzki, R. P., \& Najarro, F. 2006, ApJ, 653, L113 
\bibitem[Kroupa (2002)] {kroupa02} Kroupa, P. 2002, Science, 295, 82
\bibitem[Larson (1981)] {larson81} Larson, R. B. 1981, MNRAS, 194, 809 
\bibitem[Lin (2007)] {lin07} Lin, D. N. C., \& Murray, S. D. 2007, ApJ, accepted, (astro-ph/0703807)
\bibitem[Matzner (2000)] {matzner00} Matzner, C. D., \& McKee, C. F.  2000, ApJ, 545, 364
\bibitem[Miller (1979)] {miller79} Miller, G. E., \& Scalo, J. M. 1979, ApJS, 41, 513 
\bibitem[Mouri (2002)] {mouri02} Mouri, H., \& Taniguchi, Y.  2002, ApJ, 580, 844 
\bibitem[Nakano (1966)] {nakano66} Nakano, T. 1966, Prog. Theo. Phys., 36, 515 
\bibitem[Padoan  (1997)] {padoan07} Padoan, P., Nordlund, \AA., \& Jones, B. J. T. 1997, MNRAS, 288, 145 
\bibitem[Padoan  (2002)] {padoan02} Padoan, P., \& Nordlund, \AA. 2002, ApJ, 576, 870 
\bibitem[Portegies Zwart (2007)] {portegies07} Portegies Zwart, S., Gaburov, E., Chen, H.-C., \& Atakan G\"{u}rkan, M. 2007, MNRAS, accepted, (astro-ph/0702693) 
\bibitem[Rieke (1993)] {rieke93} Rieke, G. H., Loken, K., Rieke, M. J., \& Tamblyn, P. 1993, ApJ, 412, 99  
\bibitem[Salpeter (1955)] {salpeter55} Salpeter, E. E. 1955, ApJ, 121, 161 
\bibitem[Shadmehri (2004)] {shadmehri04} Shadmehri, M. 2004, MNRAS, 354, 373 
\bibitem[Silk (1979)] {silk79} Silk, J., \& Takahashi, T. 1979, ApJ, 229, 242 
\bibitem[Sternberg (1998)] {sternberg98} Sternberg, A. 1998, ApJ, 506, 721 
\bibitem[Stolte (2002)] {stolte02} Stolte, A., Grebel, E. K., Brandner, W., \& Figer, D. F. 2002, A\&A, 394, 459 
\bibitem[Stolte (2005)] {stolte05} Stolte, A., Brandner, W., Grebel, E. K., Lenzen, R., \& Lagrange, A.-M. 2005, ApJ, 628, L113
\bibitem[Vesperini (2001)] {vesperini} Vesperini, E., \& Heggie, D. C. 1997, MNRAS, 289, 898 
\bibitem[Whitworth (2001)] {whitworth} Whitworth, A. P., \& Ward-Thompson, D. 2001, ApJ, 547, 317
 
\end{thebibliography}
\end{document}